\documentstyle[12pt,epsfig,epsf]{article}
\title{{Effects of R-parity Violation on the Charged Higgs Boson Decays}}
\author{Yi Ping Song, Chong Sheng Li\footnote{csli@pku.edu.cn}\,,\ Qiang Li and Jian Jun Liu\\
{\small Department of Physics, Peking University, Beijing 100871,
China} }

\begin{document}
\pagestyle{plain} \setcounter{page}{1} \baselineskip=0.3in

\maketitle

\vspace{.2in}
\begin{footnotesize}
\begin{center}\begin{minipage}{5in}
\baselineskip=0.25in
\begin{center} ABSTRACT \end{center}
We calculate one-loop R-parity-violating couplings corrections to
the processes $H^-\rightarrow \tau\bar{\nu_\tau}$ and
$H^-\rightarrow b\bar{t}$. We find that the corrections to the
$H^-\rightarrow \tau\bar{\nu_\tau}$ decay mode are generally about
$0.1\%$, and can be negligible. But the corrections to the
$H^-\rightarrow b\bar{t}$ decay mode can reach a few percent for
the favored parameters.

\end{minipage}\end{center}
\vspace{3.5cm}

\end{footnotesize}



 \eject \baselineskip=0.3in
\begin{center} {\Large 1. Introduction}\end{center}

The minimal supersymmeytic standard model(MSSM) takes the minimal
Higgs structure of two doublets\cite{MSSMHiggsForm}, which
predicts the existence of three neutral and two charged Higgs
bosons $h^0,H^0,A^0,$ and $H^{\pm}$. When the Higgs boson of the
Standard Model(SM) has a mass below 130-140 Gev and the $h^0$ of
the MSSM are in the decoupling limit (which means that $H^\pm$ is
too heavy anyway to be possibly produced), the lightest neutral
Higgs boson may be difficult to be distinguished from the neutral
Higgs boson of the standard model(SM). But charged Higgs bosons
carry a distinctive signature of the Higgs sector in the MSSM.
Therefore, the search for charged Higgs bosons is very important
for probing the Higgs sector of the MSSM, and will be one of the
prime objectives of the CERN Large Hadron Collider(LHC).

Current bounds on charged Higgs mass can be obtained at the
Tevatron, by studying the top decay $t\to bH^+$, which already
eliminates some region of parameter space \cite{LHC}, whereas the
combined LEP experiments gives a low bounds approximately
$m_{H^+}>78.6$GeV at $95\%$CL\cite{LEP}. In the MSSM, we have
$m_{H^{\pm}}\ge 120$ GeV from the mass bounds from LEP--II for the
neutral pseudoscalar $A^0$ of the MSSM ($m_{A^0}\ge 90.5$
GeV)\cite{H+mass}.

Decays of a charged Higgs boson have been studied in the
literature\cite{H+decay}, which have shown that the dominate decay
modes of the charged Higgs boson for large $\tan\beta$ are
$H^\pm\rightarrow tb$ and $\tau\nu_\tau$, while $H^\pm\rightarrow
tb$, $\tau\nu_\tau$ and $Wh$ for small $\tan\beta$. For example,
for $m_{H^+}$ = 250 GeV, we have $Br(H^+ \to t\bar b) = 0.90$ and
$Br(H^+ \to \tau^+\nu_\tau) = 0.06$ for $\tan\beta = 5$, and
$Br(H^+ \to t\bar b) = 0.64$ and $Br(H^+ \to \tau^+\nu_\tau) =
0.36$ for $\tan\beta = 30$. Moreover, if charged Higgs boson mass
$m_{H^\pm}$ is very heavy, the decay of $H^\pm$ into
$\tilde{{\chi}^\pm_i}\tilde{{\chi}^0_j}$ are also
important\cite{H+chi}.

For all these decay channels, the one-loop Electroweak, QCD and
SUSY-QCD corrections have been studied in detail in the previous
literatures, for example see \cite{radiativec}. However, those
one-loop effects were studied only in the MSSM with the discrete
multiplicative symmetry of R-parity$\cite{RP}$, and without
R-parity, the effects of one-loop R-parity violating couplings on
the decays of charged Higgs boson have not reported in the
literatures so far. In this paper, we try to fill this gap and
present the calculation of the R-parity violating effects to the
process $H^-\rightarrow b\bar{t}$ and $H^-\rightarrow
\tau\bar{\nu_\tau}$, which arise from the virtual effects of
R-parity Violating couplings. The most general superpotential of
the MSSM consistent with the $SU(3)\times SU(2)\times U(1)$
symmetry and supersymmetry contains R-violating interactions,
which are given by\cite{RPTerm}
\begin{eqnarray}
{\cal W}_{\not
R}=\frac{1}{2}\lambda_{ijk}L_iL_jE_k^c+\lambda'_{ijk}\delta^{\alpha\beta}L_iQ_{j\alpha}D^c_{k\beta}
+\frac{1}{2}\lambda^{''}_{ijk}\varepsilon^{\alpha\beta\gamma}U^c_{i\alpha}D^c_{j\beta}D^c_{k\gamma}+\mu_iL_iH_2.
\end{eqnarray}
\noindent Here $L_i (Q_i)$ and $E_i (U_i, D_i)$ are, respectively,
the left-handed lepton (quark) doublet and right-handed lepton
(quark) singlet chiral superfields, and  $H_{1,2}$ are the Higgs
chiral superfields. The indices $i, j, k$ denote generations and
$\alpha$ $\beta$ and $\gamma$ are the color indices, and the
superscript $c$ denotes charge conjugation. The $\lambda$ and
$\lambda'$ are the coupling constants of L(lepton
number)-violating interactions and $\lambda^{''}$ those of
B(baryon number)-violating interactions. The non-observation (so
far) of the proton decay imposes very strong constraints on the
product of L-violating and B-violating couplings. It is thus
conventionally assumed in the phenomenological studies that only
one type of these interactions (either L- or B-violating) exists.
Some constraints on these R-parity violating couplings have been
obtained from various analysis of their phenomenological
implications based on experiment\cite{rpc}.

\begin{center} {\Large 2. Calculations}\end{center}

The tree-level amplitudes of the two decay modes of charged Higgs
boson, as shown in Fig.1(a), are given by
\begin{eqnarray}
M^{(0)}_1=\frac{ie\tan\beta m_\tau}{\sqrt{2}m_Ws_W}\bar{u}_\tau
P_Lv_{\nu}
\end{eqnarray}
\noindent for  $H^-\rightarrow \tau\bar{\nu_\tau}$, and
\begin{eqnarray}
M^{(0)}_2=\frac{ie}{\sqrt{2}m_Ws_W}\bar{u}_b(m_b\tan\beta
P_L+m_t/\tan\beta P_R)v_t
\end{eqnarray}
\noindent for  $H^-\rightarrow b\bar{t}$, where
$s_W\equiv\sin\theta_W=1-m_W^2/m_Z^2$,
$P_{R,L}\equiv(1\pm\gamma_5)/2$.

The above amplitudes lead to the tree-level decay width of the
form
\begin{eqnarray}
\Gamma^{(0)}_s =\frac{{\overline{\sum}}{|M^{(0)}|}^2\lambda^{1/2}
(m_{H^-}^2, a^2_s, b^2_s)}{16\pi m_{H^-}^3},
\end{eqnarray}
where ${\overline{\sum}}{|M^{(0)}|}^2$ is the squared matrix
element, which has been summed the colors and spins of the out
going particles, $\lambda(x,y,z)=(x-y-z)^2-4yz$, and s=(1,2)
corresponds to the decays into $\tau\bar{\nu_\tau}, b\bar{t}$,
with $a_1=m_\tau$, $b_1=0$, and $a_2=m_b$, $b_2=m_t$,
respectively.

Feynman diagrams contributing to the R-parity violating
corrections to $H^-\rightarrow \tau\bar{\nu_\tau}, b\bar{t}$  are
shown in Fig.1(b)--(c).

We carried out the calculation in the t'Hooft-Feynman gauge and
used dimensional reduction, which preserves supersymmetry, for
regularization of the ultraviolet divergences in the virtual loop
corrections using the on-mass-shell renormalization
scheme\cite{on-mass}, in which the fine-structure constant
$\alpha_{ew}$ and physical masses are chosen to be the
renormalized parameters, and finite parts of the counterterms are
fixed by the renormalization conditions. The coupling constant $g$
is related to the input parameters $e$, $m_W,$ and $m_Z$ via $g^2=
e^2/s^2_W$ and $s^2_w=1-m^2_w/m^2_Z$.

The relevant renormalization constants in the calculations of the
processes $H^-\rightarrow \tau\bar{\nu_\tau}, b\bar{t}$ are
defined as
\begin{eqnarray}
&& m_{f0}=m_{f} +\delta m_{f}, \ \ (f=\tau,t,b) \\
&& \psi_{f0}=(1+\delta Z_{fL})^{\frac{1}{2}}\psi_{fL}+(1+\delta
 Z_{fR})^{\frac{1}{2}}\psi_{fR}, \ \ (f=\tau,\nu,t,b)\\
&& \tan\beta_0=(1+\delta Z_\beta)\tan\beta.
\end{eqnarray}
 For $\delta Z_\beta$, we use
the on-shell fixing condition\cite{tbeta}
\begin{eqnarray}
 \rm{Im} \{ \hat{\Pi}_{A^0Z^0}(m^2_{A^0}) \}=0,
 \end{eqnarray} where
$\hat{\Pi}_{A^0Z^0}(m_{A^0}^2)$ is the renormalized self--energy
for the mixing of the pseudoscalar Higgs boson $A^0$ and the $Z^0$
boson, then we have
\begin{eqnarray}
\delta Z_{\beta}=\rm{Im} \{ \Pi_{A^0Z^0}(m^2_{A^0})
\}/(m_{Z^0}\sin 2\beta).
\end{eqnarray}
Apparently, there are no R-parity violating contributions to
$\Pi_{A^0Z^0}$ in our case, which leads to $\delta Z_{\beta}=0$.

Taking into account the R-parity violating corrections, the
renormalized amplitudes for $H^-\rightarrow \tau\bar{\nu_\tau},
b\bar{t}$ can be written as
\begin{equation}
M^{ren}_s=M^{(0)}_s +\delta M^{(v)}_s +\delta M^{(c)}_s,
\end{equation}
where $\delta M^{(v)}_s$ and $\delta M^{(c)}_s$ are the vertex
corrections and  the counterterms, respectively. The calculations
of the vertex corrections from Fig.1(b)-1(c) result in
\begin{eqnarray}
&&\delta
M^{(v)}_{s=1}=\frac{ig}{16\sqrt{2}\pi^2m_W}(\lambda^\prime_{333})^2\bigg\{\sum_{m=1}^2\bar{u}_\tau
\big\{[-\gamma_\mu(m^2_t/\tan\beta+m^2_b\tan\beta)C_\mu\nonumber\\
&&
\hspace{1.0cm}+(\not{\!p}_1m^2_b\tan\beta-\not{\!p}_2m^2_t/\tan\beta)C_0](m^2_\tau,m^2_{H^-},0,m^2_{\tilde{b}_m},
m^2_t,m^2_b)\big\}P_Lv_\nu\nonumber\\
&&\hspace{1.0cm}+\sum^{2}_{m,n=1}G_{mn}\bar{u}_\tau[(\gamma_\mu
P_L)C_\mu(m^2_\tau,m^2_{H^-},0,m^2_b,m^2_{\tilde{t}_m},m^2_{\tilde{b}_n})]v_\nu\bigg\},
\\
&&\delta
M^{(v)}_{s=2}=\frac{ig}{16\sqrt{2}\pi^2m_w}(\lambda^{\prime\prime}_{332})^2\bigg\{\sum_{m=1}^22(R^{\tilde{s}}_{m2})^2\bar{u}_b
[-\gamma_\mu(m_tm_b\cot\beta+m_tm_b\tan\beta)C_{\mu}\nonumber\\
&&
\hspace{1.0cm}+(\not{\!p}_1m_tm_b\cot\beta-\not{\!p}_2m_tm_b\tan\beta)C_0](m^2_b,m^2_{H^-},m^2_t,m^2_{\tilde{s}_m},m^2_t,m^2_b)P_Rv_{t}\nonumber\\
&&\hspace{1.0cm}+\sum^{2}_{m,n=1}R^{\tilde{t}}_{m2}R^{\tilde{b}}_{n2}G_{mn}\bar{u}_b[\gamma_{\mu}C_{\mu}]
(m^2_b,m^2_{H^-},m^2_t,0,m^2_{\tilde{t}_m},m^2_{\tilde{b}_n})P_Rv_{t}\bigg\},
\end{eqnarray}
with
\begin{eqnarray}
&&\hspace{-1.2cm}G_{mn}=-m_b(\mu+A_b\tan\beta)\tan\beta
R^{\tilde{b}}_{m2}R^{\tilde{t}}_{n1}-m_t(A_t+\mu\tan\beta)R^{\tilde{t}}_{n2}R^{\tilde{b}}_{m1}
\nonumber\\&&\hspace{-0.9cm}+m_tm_b(1+\tan^2\beta)R^{\tilde{t}}_{n2}R^{\tilde{b}}_{m2}
+\left\{[\tan\beta(m^2_w\sin\,2\beta-m^2_{b}\tan\beta
)-m^2_{t}]R^{\tilde{b}}_{m1}R^{\tilde{t}}_{n1}\right\},
\end{eqnarray}
where $C_0,C_\mu$ are the three-point Feynman
integrals\cite{denner},
$A_{t,b}$ are soft SUSY-breaking parameters, $\mu$ is the higgsino
mass parameter, $m_{\tilde{t}(\tilde{b},\tilde{s})_{1,2}}$ are the
stop(sbottom, sstrange) masses, and $R^{\tilde t({\tilde b})}$ are
$2\times 2$ matrix, which are defined to transform the
stop(sbottom) current eigenstates to the mass eigenstates.

The counterterms can be expressed as
\begin{eqnarray}
\delta M_{s=1}^{(c)}&=&\frac{ig\tan\beta
m_{\tau}}{\sqrt{2}m_w}\left(\frac{\delta
m_{\tau}}{m_{\tau}}+\frac{1}{2}\delta Z_{\tau R}+\frac{1}{2}\delta
Z_{\nu L}\right)\bar{u}_{\tau}P_Lv_{\nu},\\
\delta
M_{s=2}^{(c)}&=&\frac{ig}{\sqrt{2}m_w}\left[m_b\tan\beta\left(\frac{\delta
m_b}{m_b}+\frac{1}{2}\delta Z_{bR}+\frac{1}{2}\delta
Z_{tL}\right)\bar{u}_bP_Lv_{\nu}\right.\nonumber\\
&&\left.\ \ \ \ \ \ \ \ \ +m_t\cot\beta\left(\frac{\delta
m_t}{m_t}+\frac{1}{2}\delta Z_{bL}+\frac{1}{2}\delta
Z_{tR}\right)\bar{u}_bP_Rv_{\nu}\right].
\end{eqnarray}
  Calculating the self-energy diagrams in Fig.2, we can get the
explicit expressions of the renormalization constants as follows:
\begin{eqnarray}
\frac{\delta
m_{\tau}}{m_{\tau}}&=&\frac{1}{32\pi^2}(\lambda'_{333})^2\times\nonumber\\&&\sum^{2}_{m=1}
\left\{(R^{\tilde{t}}_{m1})^2[B_1+B_0](m^2_{\tau},m^2_{\tilde{t}_m},m^2_{b})+
(R^{\tilde{b}}_{m2})^2[B_1+B_0](m^2_{\tau},m^2_{\tilde{b}_m},m^2_t)\right\},\nonumber\\\\
\delta Z^{\tau}_R &=&
-\frac{m^2_{\tau}}{16\pi^2}(\lambda'_{333})^2\times\nonumber\\
&&\sum^{2}_{m=1}
\left\{(R^{\tilde{t}}_{m1})^2[B'_1+B'_0](m^2_{\tau},m^2_{\tilde{t}_m},m^2_b)+
(R^{\tilde{b}}_{m2})^2[B'_1+B'_0](m^2_{\tau},m^2_{\tilde{b}_m},m^2_t)\right\},\nonumber\\
  \\
\delta Z^{\nu}_L &=&
\frac{-1}{16\pi^2}(\lambda'_{333})^2\times\nonumber\\
&&\sum^{2}_{m=1}
\left\{(R^{\tilde{b}}_{m1})^2[B_1+B_0](0,m^2_{\tilde{b}_m},m^2_b)+
(R^{\tilde{b}}_{m2})^2[B_1+B_0](0,m^2_{\tilde{b}_m},m^2_b)\right\},\nonumber\\
  \\
\frac{\delta
m_b}{m_b}&=&\frac{1}{16\pi^2}(\lambda^{''}_{332})^2\times\nonumber\\
&&\sum^{2}_{m=1}\left\{(R^{\tilde{s}}_{m2})^2[B_1+B_0](m^2_b,m^2_{\tilde{s}_m},m^2_{t})
\right.\left.+\frac{1}{4}(R^{\tilde{t}}_{m2})^2[B_1+B_0](m^2_b,m^2_{\tilde{t}_m},m^2_{s})\right\},\nonumber\\
  \\
\delta
Z_{bL}&=&-\frac{m^2_b}{8\pi^2}(\lambda^{''}_{332})^2\times\nonumber\\
&&\sum^{2}_{m=1}\left\{(R^{\tilde{s}}_{m2})^2[B'_1+B'_0](m^2_b,m^2_{\tilde{s}_m},m^2_{t})\right.
\left.+\frac{1}{4}(R^{\tilde{t}}_{m2})^2[B'_1+B'_0](m^2_b,m^2_{\tilde{t}_m},m^2_{s})\right\},\nonumber\\
  \\
\delta
Z_{bR}&=&-\frac{1}{8\pi^2}(\lambda^{''}_{332})^2\times\nonumber\\
&&\sum^{2}_{m=1}\left\{(R^{\tilde{s}}_{m2})^2[B_1+B_0](m^2_b,m^2_{\tilde{s}_m},m^2_{t})\right.
\left.+\frac{1}{4}(R^{\tilde{t}}_{m2})^2[B_1+B_0](m^2_b,m^2_{\tilde{t}_m},m^2_{s})\right\}\nonumber\\
&&-\frac{m^2_b}{8\pi^2}(\lambda^{''}_{332})^2\times\nonumber\\
&&\sum^{2}_{m=1}\left\{(R^{\tilde{s}}_{m2})^2[B'_1+B'_0](m^2_b,m^2_{\tilde{s}_m},m^2_{t})\right.
\left.+\frac{1}{4}(R^{\tilde{t}}_{m2})^2[B'_1+B'_0](m^2_b,m^2_{\tilde{t}_m},m^2_{s})\right\},\nonumber\\
  \\
\frac{\delta
m_t}{m_t}&=&\frac{1}{16\pi^2}\sum^{2}_{m=1}(\lambda''_{332})^2(R^{\tilde{s}}_{m2})^2
[B_1+B_0](m^2_t,m^2_{\tilde{s}_m},m^2_{b}),\\
\delta
Z_{tL}&=&-\frac{m^2_t}{8\pi^2}\sum^{2}_{m=1}(\lambda''_{332})^2(R^{\tilde{s}}_{m2})^2
[B'_1+B'_0](m^2_t,m^2_{\tilde{s}_m},m^2_{b}),\\
\delta
Z_{tR}&=&-\frac{1}{8\pi^2}\sum^{2}_{m=1}(\lambda''_{332})^2(R^{\tilde{s}}_{m2})^2
[B_1+B_0](m^2_t,m^2_{\tilde{s}_m},m^2_{b})\nonumber\\
&&-\frac{m^2_t}{8\pi^2}\sum^{2}_{m=1}(\lambda''_{332})^2(R^{\tilde{s}}_{m2})^2
[B'_1+B'_0](m^2_t,m^2_{\tilde{s}_m},m^2_{b}),
\end{eqnarray}
where $B_0,B_1$ are the two-point Feynman integrals\cite{denner},
and $B^\prime_1(p^2,m^2_1,m^2_2)$ $=$ $\partial B_1/\partial p^2$,
$B^\prime_0(p^2,m^2_1,m^2_2)$ $=$ $\partial B_0/\partial p^2$.

Above we have shown the expressions of the contributions from the
couplings $\lambda^{'}_{333}$ and $\lambda^{''}_{332}$, while the
ones from the couplings $\lambda^{'}_{331}$, $\lambda^{'}_{332}$
and $\lambda^{''}_{331}$ are similar, and can be obtained
straightforwardly by substituting the corresponding particle
masses.

Finally, the renormalized decay width is then given by
\begin{eqnarray}
\Gamma_s=\Gamma^{(0)}_s +\delta \Gamma^{(v)}_s +\delta
\Gamma^{(c)}_s
\end{eqnarray}
with
\begin{eqnarray}
\delta \Gamma^{(h)}_s =\frac{\lambda^{1/2}(m_{H^-}^2, a_{s}^2,
b_{s}^2)}{8\pi m_{H^-}^3} {\rm Re} \{{\sum}{M^{(0)\ast}_s \delta
M^{(h)}_s\}} \ \ \ \ \ (h=v,c).
\end{eqnarray}
\vspace{0.1cm}\\
\begin{center}{\Large 3. Numerical results and conclusions}\end{center}

We now present some numerical results for the  R-parity violating
effects on the processes $H^-\rightarrow b\bar{t}$ and
$H^-\rightarrow \tau\bar{\nu}_\tau$. The SM input parameters in
our calculations were taken to be $\alpha_{ew}(m_Z)=1/128.8$,
$m_W=80.419$GeV and $m_Z=91.1882$GeV\cite{SM}, and
$m_t=178.0$GeV\cite{top} and $m_b(m_b)=4.25$GeV\cite{mb}.

In our calculation, we take the running mass $m_b(Q)$ and $m_t(Q)$
evaluated by the NLO formula \cite{runningmb}:
\begin{eqnarray}
&&m_b(Q)=U_6(Q,m_t)U_5(m_t,m_b)m_b(m_b),\nonumber\\
&&m_t(Q)=U_6(Q,m_t)m_t(m_t).
\end{eqnarray}
The evolution factor $U_f$ is
\begin{eqnarray}
U_f(Q_2,Q_1)=\bigg(\frac{\alpha_s(Q_2)}{\alpha_s(Q_1)}\bigg)^{d^{(f)}}
\bigg[1+\frac{\alpha_s(Q_1)-\alpha_s(Q_2)}{4\pi}J^{(f)}\bigg], \nonumber \\
d^{(f)}=\frac{12}{33-2f}, \hspace{1.0cm}
J^{(f)}=-\frac{8982-504f+40f^2}{3(33-2f)^2}.
\end{eqnarray}
In addition, in order to improve the perturbation calculations,
especially for large $\tan\beta$, we made the following
replacement in the tree-level couplings \cite{runningmb}:
\begin{eqnarray}
&& m_b(Q) \ \ \rightarrow \ \ \frac{m_b(Q)}{1+\Delta m_b},
\label{deltamb}
\\
&& \Delta m_b=\frac{2\alpha_s}{3\pi}M_{\tilde{g}}\mu\tan\beta
I(m_{\tilde{b}_1},m_{\tilde{b}_2},M_{\tilde{g}})
+\frac{h_t^2}{16\pi^2}\mu A_t\tan\beta
I(m_{\tilde{t}_1},m_{\tilde{t}_2},\mu) \nonumber \\
&& \hspace{1.0cm} -\frac{g^2}{16\pi^2}\mu M_2\tan\beta
\sum_{i=1}^2 \bigg[(R^{\tilde{t}}_{i1})^2
I(m_{\tilde{t}_i},M_2,\mu) + \frac{1}{2}(R^{\tilde{b}}_{i1})^2
I(m_{\tilde{b}_i},M_2,\mu)\bigg] \label{deltamb1}
\end{eqnarray}
with
\begin{eqnarray}
I(a,b,c)=\frac{1}{(a^2-b^2)(b^2-c^2)(a^2-c^2)}
(a^2b^2\log\frac{a^2}{b^2} +b^2c^2\log\frac{b^2}{c^2}
+c^2a^2\log\frac{c^2}{a^2}),
\end{eqnarray}
where $M_2$ is the parameter in the chargino and neutralino
matrix, and in our calculation, we always set $M_2=200$GeV.
$m_{\tilde{g}}$ is the gluino mass, which is related to $M_2$ by
$m_{\tilde{g}}=(\alpha_s(m_{\tilde{g}})/\alpha_2)M_2$\cite{Hidaka}.

The two-loop leading-log relations\cite{Higgsss} of the neutral
Higgs boson masses and mixing angles in the MSSM were used. For
$m_{H^-}$ the tree-level formula was used.

Other parameters are determined as follows:

(i) For the parameters $m^2_{\tilde{Q},\tilde{U},\tilde{D}}$ and
$A_{t,b}$ in squark mass matrices
\begin{eqnarray}
M^2_{\tilde{q}} =\left(\begin{array}{cc} M_{LL}^2 & m_q M_{LR}\\
m_q M_{RL} & M_{RR}^2 \end{array} \right)
\end{eqnarray}
with
\begin{eqnarray}
&&M_{LL}^2 =m_{\tilde{Q}}^2 +m_q^2 +m_Z^2\cos 2\beta(I_q^{3L}
-e_q\sin^2\theta_W), \nonumber
\\&& M_{RR}^2 =m_{\tilde{U},\tilde{D}}^2 +m_q^2 +m_Z^2
\cos 2\beta e_q\sin^2\theta_W, \nonumber
\\&& M_{LR} =M_{RL} =\left(\begin{array}{ll} A_t -\mu\cot\beta &
(\tilde{q} =\tilde{t}) \\ A_b -\mu\tan\beta & (\tilde{q}
=\tilde{b}) \end{array} \right),
\end{eqnarray}
to simplify the calculation we assumed
$M_{\tilde{Q}}=M_{\tilde{U}} =M_{\tilde{D}}$ and $A_t=A_b$, and we
used $m_{\tilde t_1}$, $m_{\tilde b_1}$, $A_t=A_b$ and $\mu$ as
the input parameters. We also assume $m_{{\tilde
d}_{1,2}}=m_{{\tilde s}_{1,2}}=m_{{\tilde b}_{1,2}}+500$GeV, and
$m_{{\tilde u}_{1,2}}=m_{{\tilde c}_{1,2}}=m_{{\tilde
t}_{1,2}}+500$GeV. Such assuming of the relation between the
squark masses is done merely for simplicity, and actually, our
numerical results are not sensitive to the squark masses of the
first and second generation.

(ii)According to the experimental upper bound on the couplings in
the R-parity violating interaction\cite{rpc}, we take the relevant
R-parity violating parameters as $
\lambda^{'}_{333}=\lambda^{'}_{332}= \lambda^{'}_{331}=0.3$,\ $
\lambda^{''}_{323}=-\lambda^{''}_{332}=0.9$,$ \lambda^{''}_{313}=
-\lambda^{''}_{331}=0.9$, the remainder values of $\lambda^{'}$
and $\lambda^{''}$ are set to zero. Otherwise, the numerical
results may become small because of the cancellation among the
contributions of the involved different $\lambda^{'}$ and
$\lambda^{''}$ parameters.

Fig.3 presents the dependence of the tree level decay widths on
$m_{H^-}$, where we have included the QCD and SUSY running effects
of top and bottom quark masses. From this Figure one sees that
tree level decay widths get larger with the increasing of
$m_{H^-}$.

In Fig.4 we present the relevant R-parity violating corrections to
the tree-level decay widths as the functions of $m_{H^-}$. In
general the corrections to the $\tau\bar{\nu}$ mode are negligible
small. In fact, the maximum of the corrections to $H^-\rightarrow
\tau\bar{\nu_\tau}$ are of order $0.1\%$ only. For $H^-\rightarrow
b\bar{t}$,  when $m_{H^-}>230$GeV, the corrections can be larger
than $4\%$. There are many dips and peaks on the curves, arising
from the threshold effects from the vertex corrections at the
threshold point $m_{H^-}=m_{{\tilde t}_i}+m_{{\tilde b}_j}$. For
example, as shown in Fig.4(2), at $m_{H^-}=245.9$GeV, we have
$m_{H^-}=m_{\tilde{t}_1}+m_{\tilde{b}_1}$ for $\tan\beta=40$, and
the correction to $H^-\rightarrow b\bar{t}$ can get its maximal
value of $10\%$.

Fig.5 show the dependence of the R-parity violating corrections on
$m_{\tilde{t}_1}$. In general, the corrections increase with the
decreasing of $m_{\tilde{t}_1}$. For example, when
$m_{\tilde{t}_1}=100$GeV, the correction to $H^-\rightarrow
\tau\bar{\nu_\tau}$ is $0.4\%$ for $\tan\beta=4$, while the one to
$H^-\rightarrow b\bar{t}$ is $4\%$ for $\tan\beta=40$, and when
$m_{\tilde{t}_1}=300$GeV, above corrections are both about 0.1\%.

In Fig.6 we present the R-parity violating corrections as a
function of $\tan\beta$. We find that the corrections are
relatively larger for low and high values of $\tan\beta$,
respectively, while become smaller for intermediate values of
$\tan\beta$, which is due to the fact that there are no enhanced
effects from the Yukawa couplings $H^-b\bar{t}$ and
$H^-\tilde{t}\tilde{b}$ at medium $\tan\beta$.

In conclusion, we have calculated the R-parity violating effects
on the processes $H^-\rightarrow \tau\bar{\nu_\tau}$ and
$H^-\rightarrow b\bar{t}$. These corrections arise from the
virtual effects of R-parity violating couplings. We find that the
corrections to the $H^-\rightarrow \tau\bar{\nu_\tau}$ decay mode
are generally about $0.1\%$, and can be negligible. But the
corrections to the $H^-\rightarrow b\bar{t}$ decay mode can reach
a few percent in our chosen parameter space. Compared to the
SUSY-QCD or SUSY-EW corrections, the typical values of which can
be over 10\%\cite{radiativec}, the R-parity violating effects on
the process $H^-\rightarrow b\bar{t}$ are smaller, but not
negligible in some region of the parameter space.

\section*{Acknowledgements}
\vspace{.5cm} This work was supported in part by the National
Natural Science Foundation of China and Specialized Research Fund
for the Doctoral Program of Higher Education. 
\newpage

\begin{figure}[h!]
\vspace{1.0cm} \centerline{\epsfig{file=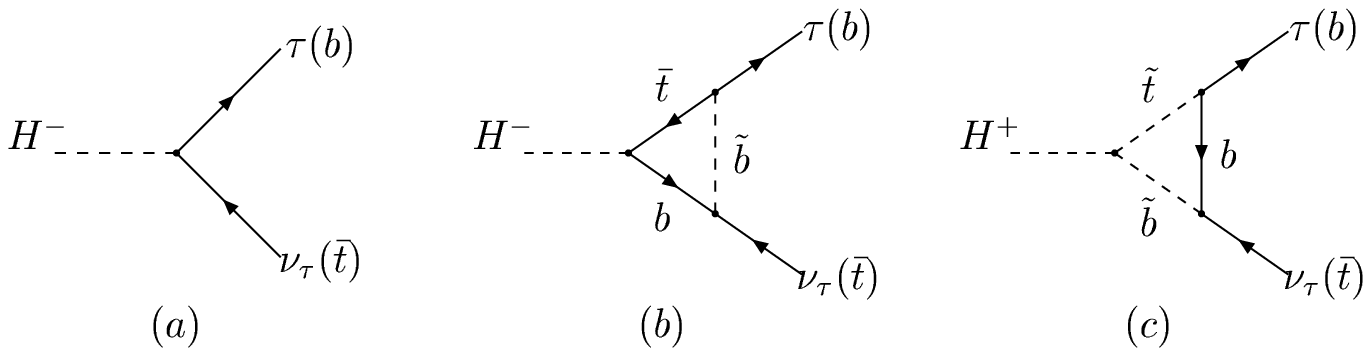,
width=420pt}}  \caption{Feynman diagrams contributing to R-parity
violating corrections to $H^-\rightarrow
\tau\nu_{\tau}(b\bar{t})$: $(a)$ tree-level diagram; $(b)-(c)$ are
one-loop vertex corrections.}
\end{figure}

\begin{figure}[h!]
\vspace{1.0cm} \centerline{\epsfig{file=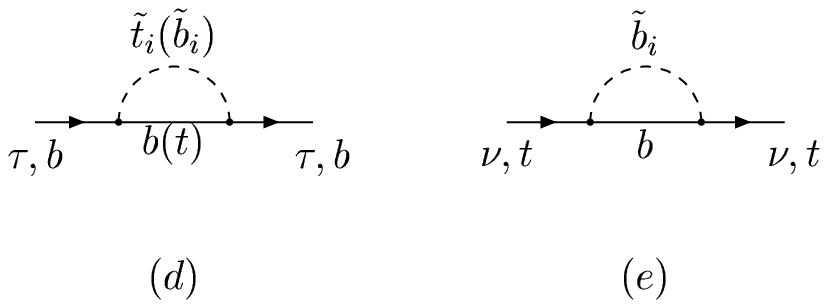,
width=320pt}}  \caption{Feynman diagrams contributing to
renormalization constants.}
\end{figure}

\begin{figure}[h!]
\vspace{1.0cm} \centerline{\epsfig{file=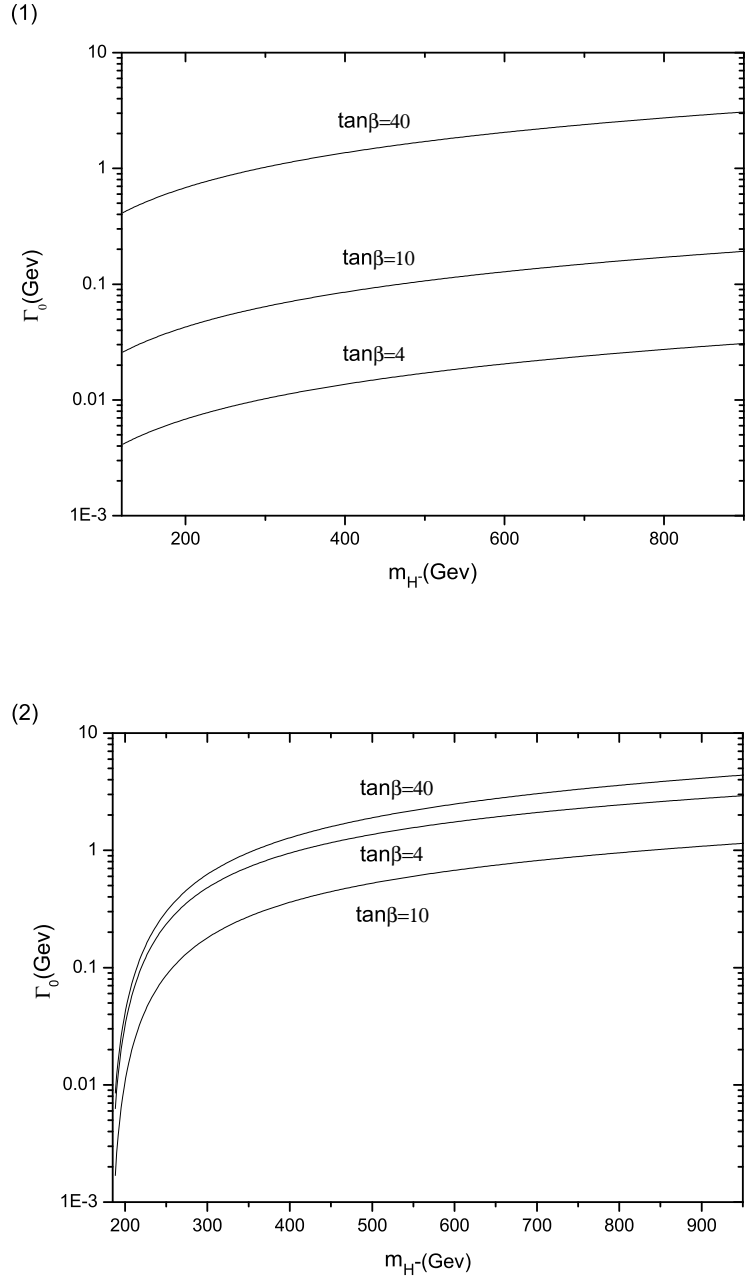, width=280pt}}
\caption{Dependence of the tree level decay widths on $m_{H^-}$
for (1) $H^-\rightarrow \tau\bar{\nu_\tau}$, assuming:
$\mu=-400$GeV, $A_t=A_b=600$GeV, and $m_{{\tilde t}_1}=100$GeV;
(2) $H^-\rightarrow b\bar{t}$, assuming: $\mu=600$GeV,
$A_t=A_b=800$GeV, and $m_{{\tilde t}_1}=100$GeV. $m_{H^-}$ runs
from 121GeV to 900GeV and 188GeV to 900GeV for (1) and (2),
respectively.}
\end{figure}

\begin{figure}[h!]
\vspace{1.0cm} \centerline{\epsfig{file=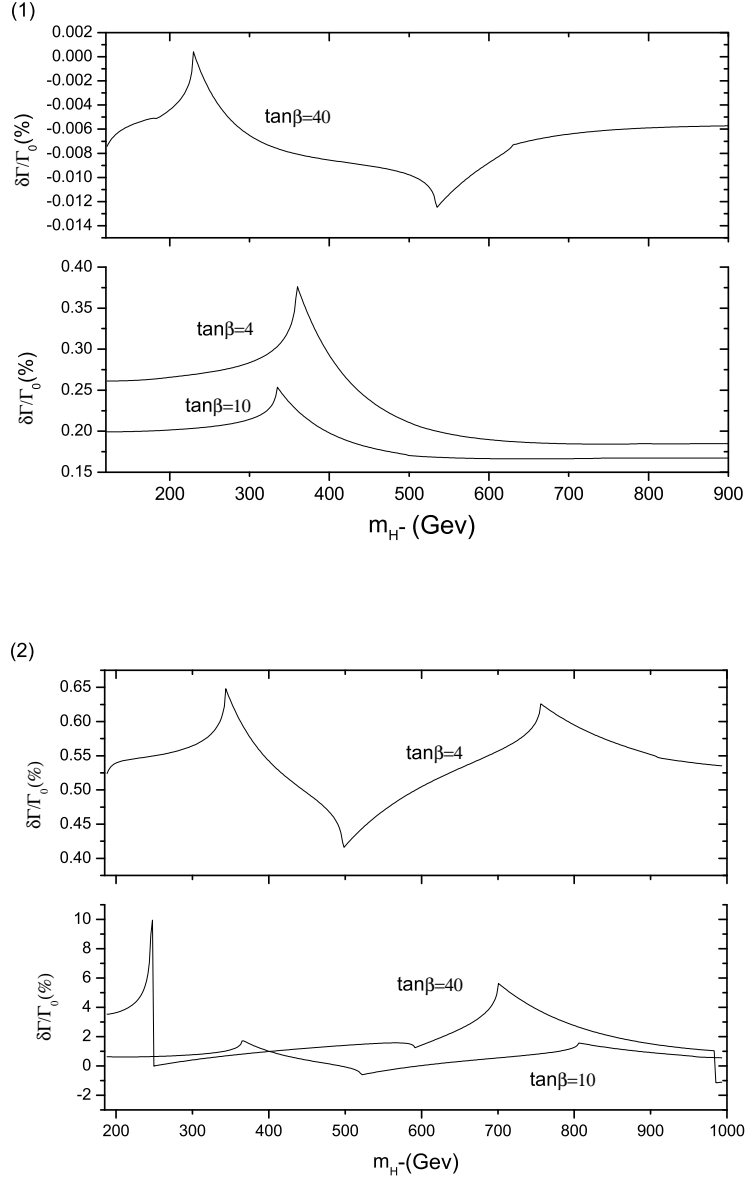, width=280pt}}
\caption{The R-parity violating corrections as functions of
$m_{H^-}$ for (1) $H^-\rightarrow \tau\bar{\nu_\tau}$, assuming:
$\mu=-400$GeV, $A_t=A_b=600$GeV, and $m_{{\tilde t}_1}=100$GeV;
(2) $H^-\rightarrow b\bar{t}$, assuming: $\mu=600$GeV,
$A_t=A_b=800$GeV, and $m_{{\tilde t}_1}=100$GeV. $m_{H^-}$ runs
from 121GeV to 900GeV and 188GeV to 900GeV for (1) and (2),
respectively.}
\end{figure}

\begin{figure}[h!]
\vspace{1.0cm} \centerline{\epsfig{file=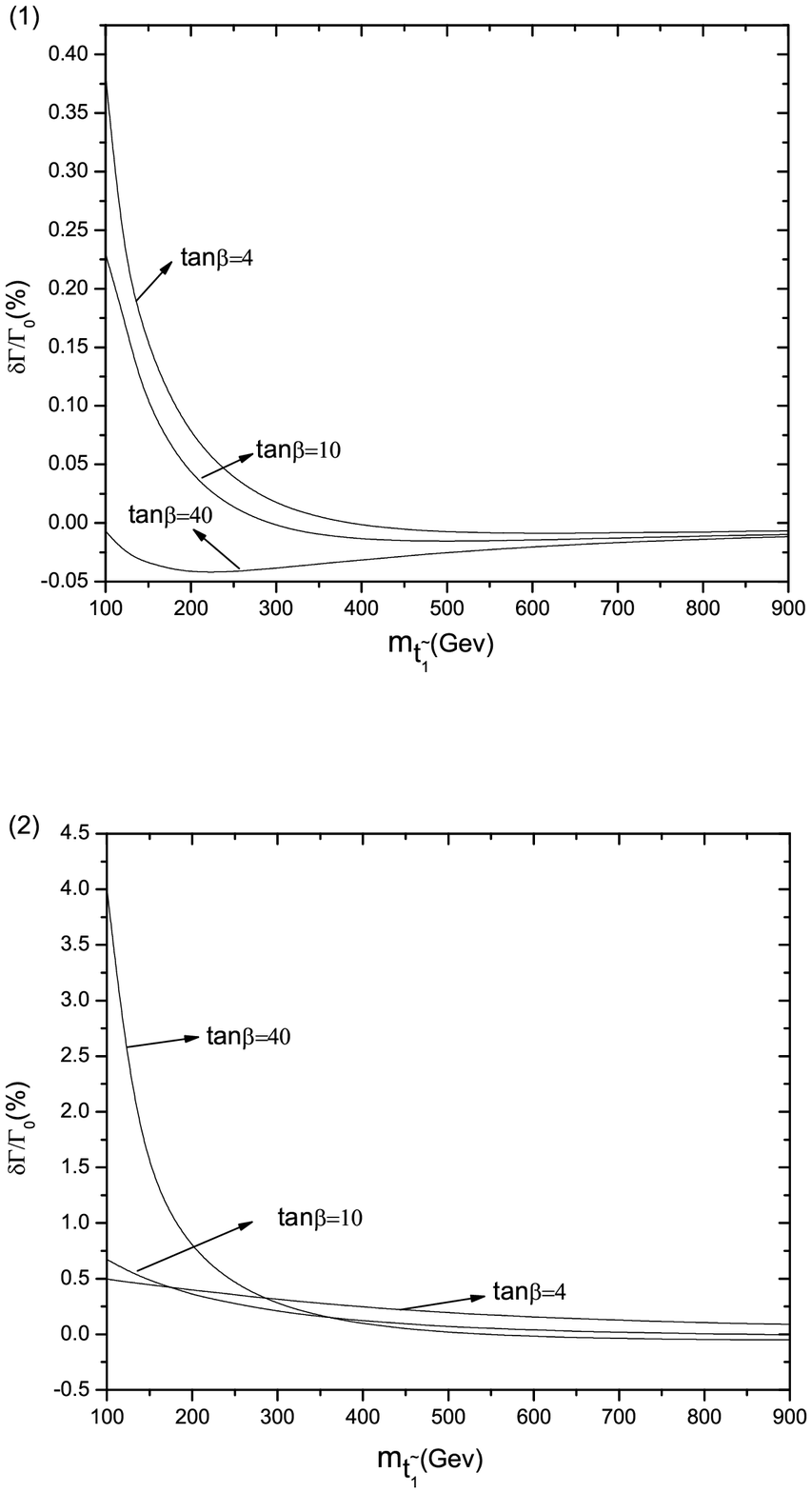, width=280pt}}
\caption{ The R-parity violating corrections as functions of
$m_{{\tilde t}_1}$ for (1) $H^-\rightarrow \tau\bar{\nu_\tau}$,
assuming: $\mu=-400$GeV, $A_t=A_b=600$GeV, and $m_{A^0}=350$GeV;
(2) $H^-\rightarrow b\bar{t}$, assuming: $\mu=600$GeV,
$A_t=A_b=800$GeV, and $m_{A^0}=200$GeV.}
\end{figure}

\begin{figure}[h!]
\vspace{1.0cm} \centerline{\epsfig{file=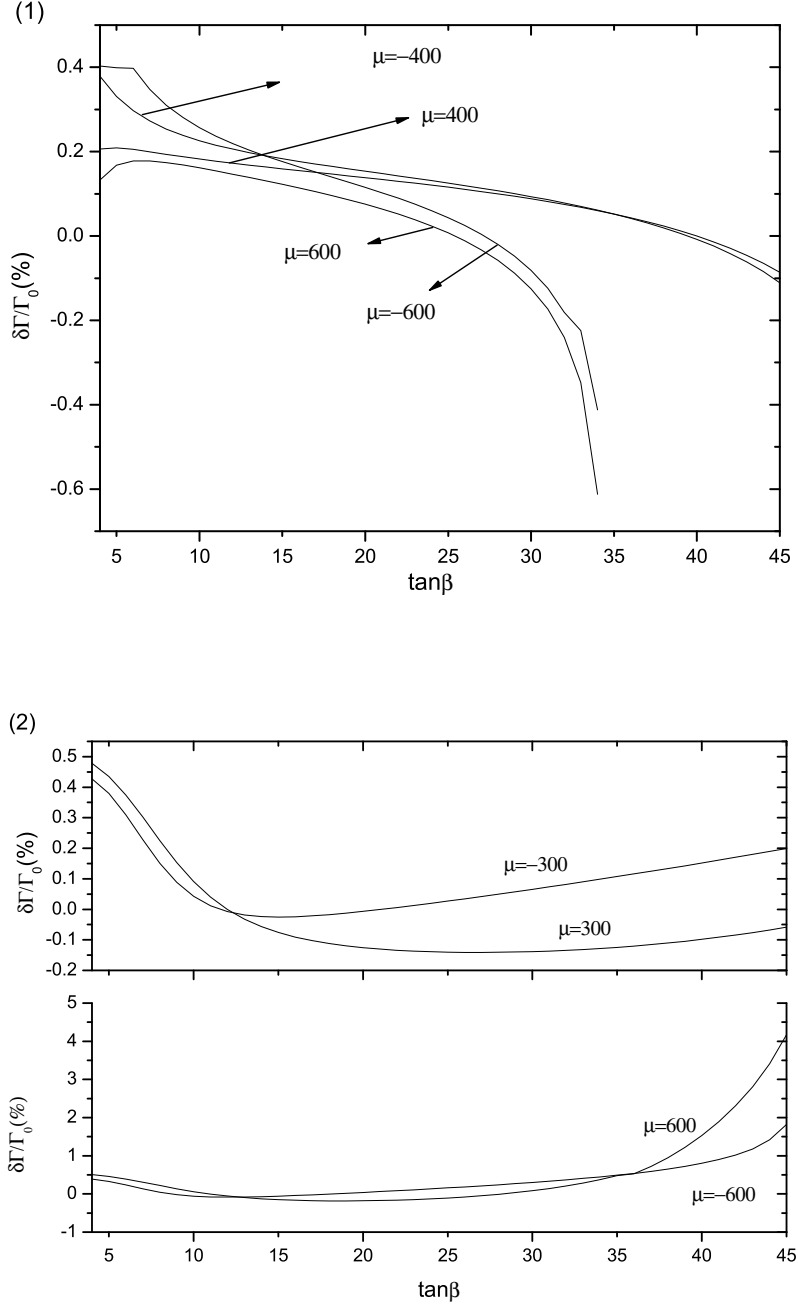, width=300pt}}
\caption{The R-parity violating corrections as functions of
$\tan\beta$ for (1) $H^-\rightarrow \tau\bar{\nu_\tau}$, assuming:
$A_t=A_b=600$GeV, $m_{A^0}=350$GeV, and $m_{{\tilde t}_1}=100$GeV;
(2) $H^-\rightarrow b\bar{t}$, assuming: $A_t=A_b=800$GeV,
$m_{A^0}=600$GeV, and $m_{{\tilde t}_1}=100$GeV.}
\end{figure}

\end{document}